\documentclass[a4paper,twocolumn,11pt]{quantumarticle}
\pdfoutput=1
\usepackage[utf8]{inputenc}
\usepackage[english]{babel}
\usepackage[T1]{fontenc}
\usepackage{amsmath,amssymb,amsfonts}
\usepackage{hyperref}
\usepackage{textcomp}
\usepackage{algorithmic}
\usepackage{tikz}
\usepackage{lipsum}
\usepackage{graphicx}

\begin{document}

\title{Spatial distribution of electric field of equal probability quantum walks based on three-level quantum system}

\author{Xiaoguang Chen}
\affiliation{Department of communications science and engineering, School of information science and engineering, Fudan University, Shanghai 200433, China}
\email{xiaoguangchen@fudan.edu.cn}
\orcid{0000-0003-1556-563x}
\affiliation{Department of communications science and engineering, School of information science and engineering, Fudan University, Shanghai 200433, China}
\maketitle

\begin{abstract}
   Based on the three-level quantum system, when it is in resonance, according to any two lattice points closest to Hamiltonian coupling, electrons transition from high energy level to low energy level and release photons; Or absorb photons and transition from low energy level to high energy level, thus obtaining the physical process of quantum walking along a straight line under the condition of equal probability. Then, the optical radiation in the quantum walk is mapped into a Gaussian pulse of the electric field, and the Maxwell’s equation is solved by the three-dimensional finite-difference time-domain method to obtain the spatial electric distributio. Finally, the physical process of quantum walking on two parallel lines is further discussed, involving some physical properties such as electromagnetic coupling or coherence, quantum state exchange and so on. The electric field coupling between two lines can be calculated by FDTD, which provides a useful tool for the design and analysis of quantum devices.
\end{abstract}

\section{Introduction}
Quantum walk is the counterpart of classical random walk in quantum mechanics. Quantum walk first appeared in Feynman's article on quantum mechanical computer \cite{1}. Today, it can be regarded as the earliest continuous quantum walk  \cite{2} model. Later, the quantum random walk proposed by Aharonov et al. \cite{3}.    can be regarded as the earliest discrete quantum walk model. Quantum walk can also be divided into atomic\cite{4,5}, ionic \cite{6}, and photonic\cite{7}. When single photons and entangled photons operate in the integrated circuit of the chip, their behavior can be coupled into harmonic oscillator or waveguide structures\cite{8,9}, which may be placed on a chip. This will greatly encourage people to study the transmission of non-classical light and the interference in the transmission process\cite{10}.  . In particular, for example, some structures can be regarded as the nearest Hamiltonian coupling, which is similar to the tightly constrained Hamiltonian, which is very famous in the field of condensed matter physics. Moreover, the integrated structure also can be allowed to do the operation of quantum logic. In addition, quantum electromagnetism, or the quantum effect in electromagnetism, or the quantization of electromagnetic field, has been given new significance by W. C. Chew et al.\cite{11,12}. At the same time, it is driven by the single photon source and measurement, the effectiveness of Bell theory, and the rapid development of nanofabrication technology.
We know that in computational theory, three state units are the most effective for classical computers\cite{13}. In the quantum world, multi-level systems are very common. We conjecture that the future quantum processor is a quantum system based on multiple energy levels. Therefore, based on the three-level quantum system, this paper will analyze the specific physical process of equal probability quantum walking. As a preliminary knowledge, we simple review the quantum walks  \cite{2,14,15,16}, the three-dimensional finite-difference time-domain method\cite{17}, and then construct the Hamiltonian operator of the open quantum system\cite{13}, mapping the trajectory of the quantum walk into the displacement of the electric field pulse. Through the three-dimensional finite-difference time-domain method, the Maxwell equation is solved, and then the spatial electromagnetic distribution is obtained.

\section{Quantum walks}
Quantum walks are typically introduced by analogy with classical random walks. The discrete quantum walks are usually called coined quantum walks [2, 25], if considering walk on an infinite line as an example, one can define an amplitude of shifting to the left adjacent site or to the right adjacent site. In this case the wave function of a walking particle, initially localized at site “0” is
\begin{equation}
|\psi(t+\Delta t)\rangle=A|-1\rangle+B|+1\rangle, |A|^2+|B|^2=1
\end{equation}
Where time interval  $\Delta t$ counts steps. One way to achieve this starting with state $|0\rangle$ is to use the state of a qubit (two-quantum system), typically referred to as “quantum coin”, to supply amplitudes for the two different directions
\begin{equation}
|\psi(t+\Delta t)\rangle=A|-1,0\rangle+B|+1,1\rangle
\end{equation}
Here the second state index denotes the basis states of the coin. In the paper, we mainly consider the discrete quantum walk and replaces the coin operator with equal probability, it will be explained in detail later.

Continuous time quantum walks do not rely on auxiliary quantum coins to propagate. It is evolution due to dynamic change of a unitary. By the processing method of graph theory and introducing a scalar incoherent parameter, and discussing a mixed quantum continuum and related classical dynamics, and solves the quantum random standard equations\cite{14,15}.

\section{Three dimensional finite difference time domain FDTD}
In 1966, Yee established a set of finite difference equations for the time-dependent Maxwell’s curl equations system  \cite{18}. These equations can be expressed in discrete form in space and time, using the second-order accurate central difference formula. The discrete positions of electric and magnetic field components in time and space are sampled. FDTD technology divides the three-dimensional problem geometry into cells to form a grid, which is composed of $N_x\times N_y\times N_z$ elements. The algorithm samples and calculates the field at discrete time points. The material parameters (dielectric constant and permeability) are distributed on the FDTD grid and are related to the field component; thus, their numbers are the same as their respective field components\cite{17}.

When the wavelength of the electromagnetic field is larger than the atomic size or the spacing of lattice points, the macro electromagnetic theory is still valid, just as expressed by classical electromagnetism\cite{19,20}. Under the same conditions, the finite-difference time-domain method is also applicable to the quantum world \cite{21}. As mentioned in reference \cite{22}, quantum information brings vitality to computational electromagnetism.

For lossless, non-dissipative and non-uniform media, we can obtain the discrete expression of quantum FDTD:
\begin{equation}
\begin{aligned}
& \hat{E}_x^{n+1}(i,j,k)=\hat{E}_x^{n}(i,j,k) \\
& +\frac{\Delta t}{\varepsilon_x(i,j,k) \Delta y}(\hat{H}_z^{n+\frac{1}{2}}(i,j,k)-\hat{H}_z^{n+\frac{1}{2}}(i,j-1,k))\\
& -\frac{\Delta t}{\varepsilon_x(i,j,k) \Delta z}(\hat{H}_y^{n+\frac{1}{2}}(i,j,k)-\hat{H}_y^{n+\frac{1}{2}}(i,j,k-1))
\end{aligned}
\end{equation}
\begin{equation}
\begin{aligned}
&\hat{H}_x^{n+\frac{1}{2}}(i,j,k) =\hat{H}_x^{n-\frac{1}{2}}(i,j,k) \\
&+\frac{\Delta t}{\mu_x(i,j,k) \Delta z}(\hat{E}_y^{n}(i,j,k+1)-\hat{E}_y^{n}(i,j,k))\\
&-\frac{\Delta t}{\mu_x(i,j,k) \Delta y}(\hat{E}_z^{n}(i,j+1,k)-\hat{E}_z^{n}
(i,j,k))
\end{aligned}
\end{equation}
Here, the classical electromagnetic field is promoted to the quantum world. Here, only the component representation of the x-axis direction of the quantum electromagnetic field in the rectangular coordinate system is given, and the components in other directions can be obtained similarly. $\Delta t$ represents the time step, $\Delta x,\Delta y,\Delta z$ which is the spatial step in the x, y, z three directions respectively.  $\varepsilon $ is dielectric constant and $\mu$ is permeability of the material and n in the equation represents the number of iterations.The "$\wedge$"in Eq.(3) and Eq.(4) represents the variables of the quantum world.

\section{Hamiltonian operator of a three-level quantum system}
For a three-level system, 1, 2 and 3 are used to represent the energy of each energy level, namely $\mathbb{E}_1$, $\mathbb{E}_2$ and $\mathbb{E}_3$. The resonant monochromatic field generated by two wave envelopes, j = 1, 2, has frequencies of and respectively. Under the rotating wave approximation (RWA)\cite{13,24,25,26}, its Hamiltonian can be written as:
\begin{equation}
\mathbb{H}=
\begin{bmatrix}
{\mathbb{E}_1}&{u_1(t)e^{i\omega_1t}}&{0}\\
{u_1(t)e^{-i\omega_1 t}}&{\mathbb{E}_2}&{u_2(t)e^{i\omega_2 t}}\\
{0}&{u_2(t)e^{-i\omega_2t}}&{\mathbb{E}_3}
\end{bmatrix}
\end{equation}
Here, $\omega_1:=\frac{\mathbb{E}_2-\mathbb{E}_1}{\hbar}$, $\omega_2:=\frac{\mathbb{E}_3-\mathbb{E}_2}{\hbar}$, the control quantity $u_j(t),j=1,2$.
The above formula can be further written as:
\begin{equation}
\mathbb{H}=D+
\begin{bmatrix}
{0}&{\Omega_1(t)}&{0}\\
{\Omega_1^*(t)}&{0}&{\Omega_2(t)}\\
{0}&{\Omega_2^*(t)}&{0}
\end{bmatrix}
\end{equation}
Here,\\ $\Omega_i(t)$=$u_i(t)e^{i\omega_it}$, $i=1,2.$ $D:$   =$diag (\mathbb{E}_1,\mathbb{E}_2,\mathbb{E}_3)$.\\That is, the energy state of the system only appears on the diagonal. This item D is called "drifting" and can be eliminated by U transformation \cite{23}.
For a three-level system, throw away the fundamental terms, there are\\
\begin{equation}
\mathbb{H}=
\begin{bmatrix}
{0}&{\Omega_1(t)}&{0}\\
{\Omega_1^*(t)}&{0}&{\Omega_2(t)}\\
{0}&{\Omega_2^*(t)}&{0}
\end{bmatrix}
\end{equation}
For complex three-level system, the system is in resonant state if and only if the following conditions are satisfied  \cite{24}:\\
\begin{equation}
\begin{cases}
\Omega_1(t)=cos(t/\sqrt3)e^{\frac{i}{\hbar}[(\mathbb{E}_2-\mathbb{E}_1)t+\varphi_1]}\\
\Omega_2(t)=sin(t/\sqrt3)e^{\frac{i}{\hbar}[(\mathbb{E}_3-\mathbb{E}_2)t+\varphi_2]}
\end{cases}
\end{equation}
Here $\varphi_1$ and  $\varphi_2$  are two arbitrary phases.
\section{Quantum walking on one line}
Suppose that in a system of three-level atoms, the electron has a transition (such as the L structure in Fig. 1), and the release of photons and is the time required for the electron to transition from the third energy level of the atom to the middle energy level, and the time required for the electron to transition from the middle (or second) energy level of the atom to the first (or lowest) energy level. According to the nearest neighbor Hamiltonian coupling principle\cite{10}, the following quantum walking can be obtained:
It is assumed that there are seven atomic lattice points on a straight line. As shown in Figure 2, the atomic lattice points are located at the coordinate position points on the one-dimensional x-axis, i.e.
$x_1=-3, x_2=-2, x_3=-1, x_4=0, x_5=1, x_6=2, x_7=3$.\\
Step 1: Assuming that at the atomic lattice point of $x_4=0$, the electron transitions from the higher energy level $(\mathbb{E}_3)$ to the middle energy level $(\mathbb{E}_2)$ , and releases the photon $\hbar\omega_1$, consuming time is $T_1$. In the resonant state, the electron transitions from the middle energy level $(\mathbb{E}_2)$ to the lower energy level $(\mathbb{E}_1)$, consuming time is $T_2$. At this moment, the atomic wave function can be written as: \\
\begin{equation}
|\psi_{x_4}\rangle=|\mathbb{E}_1\rangle
\end{equation}
At the same time, the electrons located in atomic lattice $x_3=-1$ and $x_5=1$, absorb photons $\hbar\omega_1$  and $\hbar\omega_2$  with a probability of 1/2 respectively, completing the transition from the low energy level to the high energy level. Its atomic wave function can be written as:
\begin{equation}
|\psi_{x_3}\rangle=|\psi_{x_5}\rangle=c_2|\mathbb{E}_2\rangle+c_3|\mathbb{E}_3\rangle
\end{equation}
\begin{figure}[t]
	\centering
	{ \includegraphics[scale=0.6]{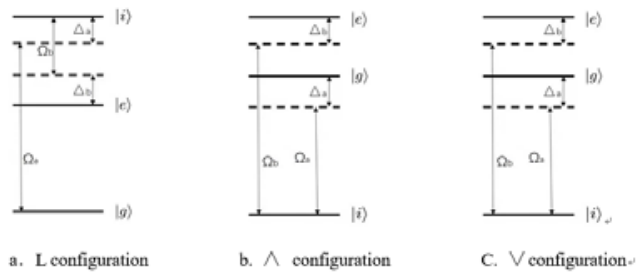}}
	\caption{Three possible configurations of atomic energy levels, when the system is in a resonant state, $\Delta a=\Delta b=0.$ $\Omega_a$ and $\Omega_b$ are the frequency of photons, respectively \cite{13}. In this paper, we use L-type configuration to analyze the problem.}
	\label{fig:figure1}
\end{figure}
\begin{figure}[t]
	\centering
	{ \includegraphics[scale=0.5]{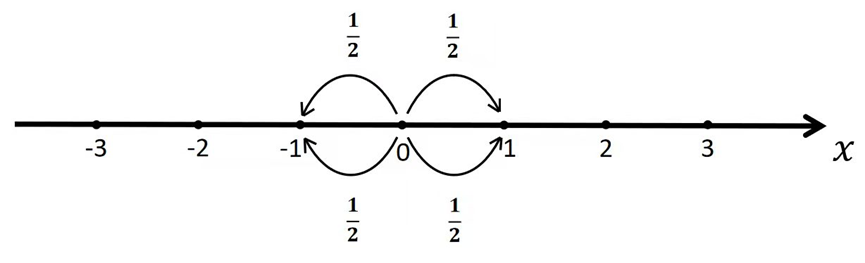}}
	\caption{The first quantum walk in a three-level atomic system, where 1/2 represents the probability of transition}
	\label{fig:figure2}
\end{figure}
Step 2: As shown in Fig.3, when the atomic lattice point is at the $x_3=-1$ and $x_5=1$, the electron transitions from the higher energy level  $(\mathbb{E}_3)$ to the middle energy level  $(\mathbb{E}_2)$ and releases the photon $\hbar\omega_1$, consuming time is $T_1$. In the resonant state, the electron transitions again from the middle energy level  $(\mathbb{E}_2)$ to the lower energy level  $(\mathbb{E}_1)$ and releases the photon $\hbar\omega_2$, consuming time is $T_2$. At the moment, the atomic wave function can be written as:
\begin{equation}
|\psi_{x_3}\rangle=|\psi_{x_5}\rangle=|\mathbb{E}_1\rangle
\end{equation}
When the atom lattice point is located in $x_2=-2$ and $x_6=2$, the electron absorbs the photon $\hbar\omega_1$ with a quarter probability and transitions from the intermediate level $(\mathbb{E}_2)$ to the higher energy level $(\mathbb{E}_3)$, which takes time $T_1$. Then, the electron absorbs photon $\hbar\omega_2$ with a quarter probability and transitions from lower energy level $(\mathbb{E}_1)$ to intermediate energy level $(\mathbb{E}_2)$, which takes time $T_2$. At this moment, the atomic wave function can be written as:
\begin{equation}
|\psi_{x_2}\rangle=|\psi_{x_6}\rangle=c_2^{'}|\mathbb{E}_2\rangle+c_3^{'}|\mathbb{E}_3\rangle
\end{equation}
\begin{figure}[t]
	\centering
	{ \includegraphics[scale=0.5]{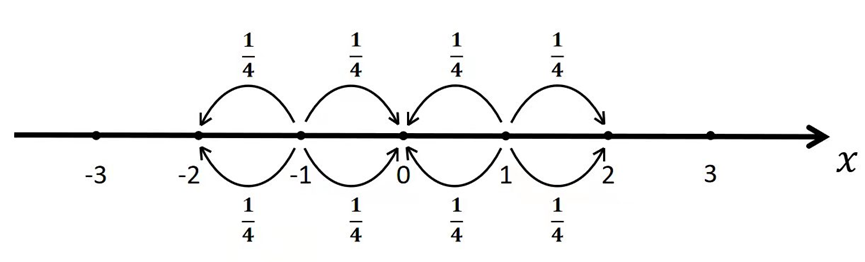}}
	\caption{The second step of quantum walking in a three-level atomic system, in where 1/4 represents the probability of transition.}
	\label{fig:figure1}
\end{figure}
The atomic lattice point located in $x_4=0$, the electron absorbs photon $\hbar\omega_1$ with a probability of $(1/4+1/4)$ and transitions from the intermediate level $\mathbb{E}_2$ to the high-energy level $\mathbb{E}_3$, which takes time $T_1$. Similarly, the electron absorbs photon $\hbar\omega_2$ with a probability of $(1/2+1/2)$ and transition from the lowest energy level $\mathbb{E}_1$ to the middle energy level $\mathbb{E}_2$, which takes time $T_2$. At this moment, the atomic wave function can be written as:
\begin{equation}
|\psi_{x_4}\rangle=c_2^{''}|\mathbb{E}_2\rangle+c_3^{''}|\mathbb{E}_3\rangle
\end{equation}
Step 3: Located in the atom lattice point $x_4=0$, electrons release the photons $\hbar\omega_1$ and $\hbar\omega_2$, and the total consuming time is $T_1+T_2$. At the same time, the electrons on the atomic lattice points $x_2=-2$  and $x_6=2$, and release photons and, and the total consuming time is still $T_1+T_2$.
At the atomic lattice points $x_3=-1$ and $x_5=1$, the electron absorbs the photon $\hbar\omega_1$ with a probability of $1/4$ and $1/8$, respectively. Similarly, the photon $\hbar\omega_2$ is absorbed with $1/4$ probability and $1/8$ probability. At last, located in $x_1=-3$ and $x_7=3$ atomic lattice points, electrons absorb photons with a probability of $1/8$. In addition, the atomic wave function at this step is omitted here and will not be repeated.\\

\begin{figure}[t]
	\centering
	{ \includegraphics[scale=0.5]{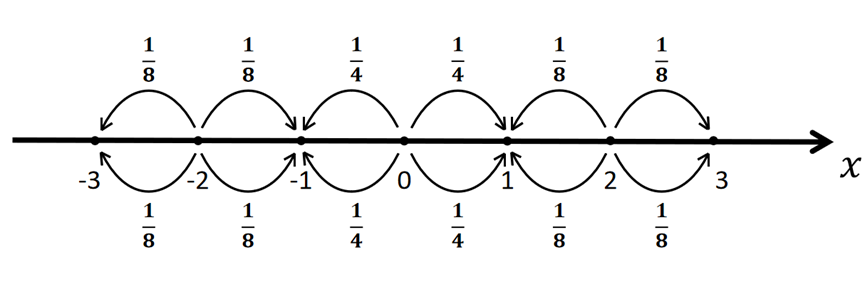}}
	\caption{The third step of quantum walking in a three-level atomic system, where $1/4$ and $1/8$ represent the probability of transition.}
	\label{fig:figure 4}
\end{figure}
\section{ Numerical calculation}
Suppose a beam of light, with a wavelength of $804nm$ and a frequency of $3.7\times10^{14}  Hz$, passes through the center of a $900nm\times90nm\times190nm$ quartz crystal, as shown in Fig.5. The first step of quantum walking is realized in the center of the crystal, the electrons transition from the high energy level to the middle energy level, releasing photon $\hbar\omega$ in the form of electromagnetic pulse, while the electrons in the surrounding atomic lattice absorb photons. We can regard this process as electric field energy storage, that is
\begin{equation}
\frac{1}{2}\varepsilon E_0^2=\hbar\omega
\end{equation}
From this, we can calculate the average electric field intensity of the energy level, further write a simple Gaussian form of electromagnetic pulse, namely:
\begin{equation}
E(t)=E_0cos(\omega_c(t-t_0))e^{-\frac{(t-t_0)^2}{\tau^2}}
\end{equation}
That is, the cosine modulated Gaussian wave packet is used to represent photons. The above formula is also the mathematical expression of the transient electric field of photons in time domain. The mathematical expression of photons in frequency domain can be obtained through Fourier transform:
\begin{equation}
E(\omega)=E_0(\frac{\tau\sqrt\pi}{2}e^{-\frac{\tau^2(\omega-\omega_0)^2}{4}}+\frac{\tau\sqrt\pi}{2}e^{-\frac{\tau^2(\omega+\omega_0)^2}{4}})
\end{equation}
This is consistent with the mathematical expression\cite{10} of the electric field when the electron is in the transition between the atomic energy levels, and also satisfies the resonance condition-Eq.(8).
\begin{figure}[t]
	\centering
	{ \includegraphics[scale=0.5]{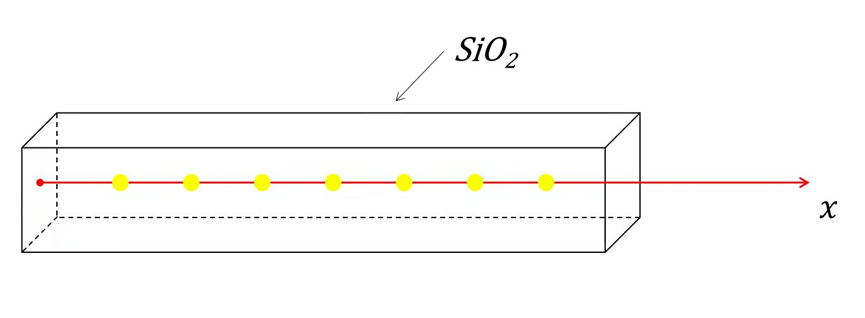}}
	\caption{A beam of red light passes through a quartz crystal, assuming that it passes through seven atomic lattice points.}
	\label{fig:figure 5}
\end{figure}

To select the time delay $t_0=4.5\tau$, and $\tau=\frac{n_c\Delta S_{max}}{2c}$,  where $\Delta S_{max}$ is the maximum value of FDTD grid step $(\Delta x, \Delta y, \Delta z)$, and $n_c$  is the number of grids per wavelength, and $c$ is the speed of light.

In addition, in FDTD simulation\cite{17}, the initial condition of the field is zero, so the electric field of the excitation source must also be zero. This can move the time Term in the Gaussian waveform by one time unit, so that the instantaneous value of the electric field at the beginning is zero. As shown in Fig. 6, when the number of iterations is $ n = 100$, a complete Gaussian electric field pulse can be obtained.
\begin{figure}[t]
	\centering
	{ \includegraphics[scale=0.4]{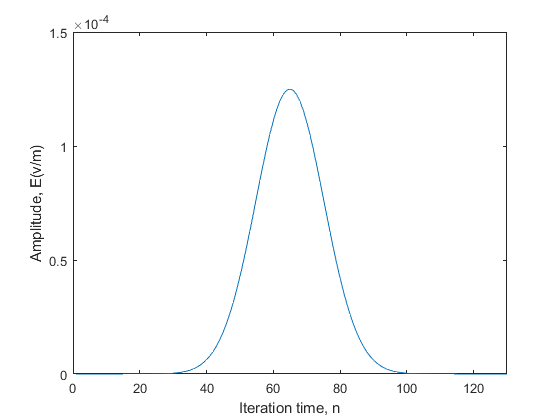}}
	\caption{The Gaussian pulse waveform.}
	\label{fig:figure 6}
\end{figure}
 Assuming that the distance between atomic lattice points is 40nm, the calculation results of the upper half of the first step of quantum walking in Fig. 2 are shown in Fig.7.
\begin{figure}[t]
	\centering
	{ \includegraphics[scale=0.24]{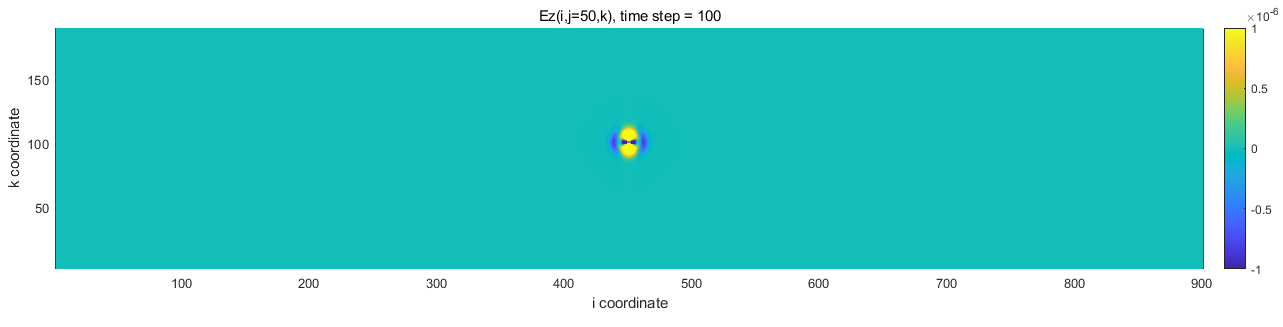}}
	\caption{The first step of quantum is the spatial electric field distribution when the electron completes the transition, in which the abscissa is in the x direction, the ordinate is in the z direction, and the unit is nm. The electric field is polarized along the z direction.}
	\label{fig:figure 7}
\end{figure}
Fig.8 is the second step of quantum walking, corresponding to the upper half of Fig.3, in which, the electric field value is only 0.707 of the corresponding value in Fig.7.
\begin{figure}[t]
	\centering
	{ \includegraphics[scale=0.24]{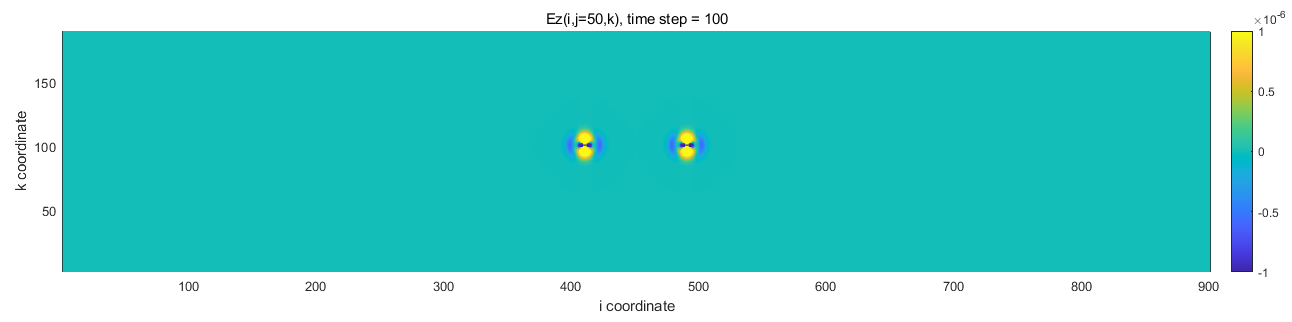}}
	\caption{The distribution of electric field in space when the electron completes the transition in the second step of quantum walking.}
	\label{fig:figure 8}
\end{figure}
\begin{figure}[t]
	\centering
	{ \includegraphics[scale=0.24]{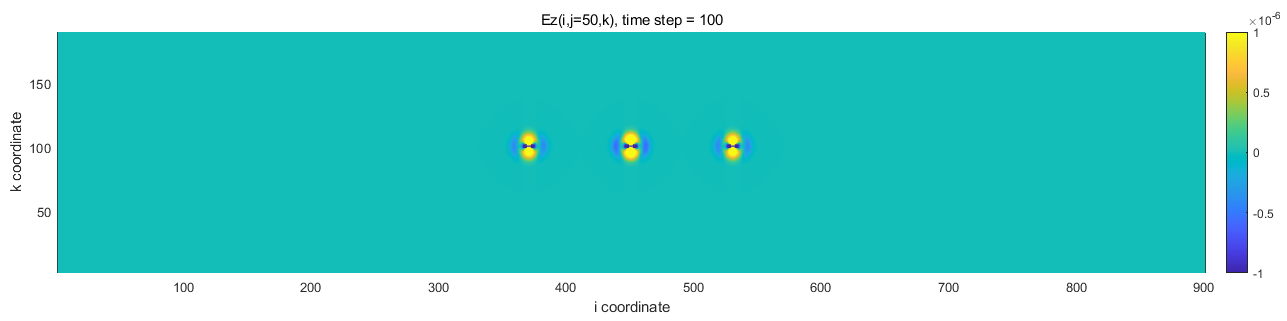}}
	\caption{The distribution of electric field in space when the electron completes the transition in the third step of quantum walking.}
	\label{fig:figure9}
\end{figure}
Fig.9 corresponds to the upper half of Fig.4. It can be seen from the above figures that the quantum electromagnetic field at this time is instantaneous and ultra-short distance. Each step of quantum walking will randomly add an excitation source.
\section{Description of quantum walking process and electric field distribution of two parallel beams of light}
As shown in Fig.10, the distances between adjacent crystal lattice points are equal. It is assumed that the initial excitation source is generated at the origin 0 of the coordinate axis, that is, the electron transitions from the third energy level $(\mathbb{E}_3)$ to the intermediate energy level $(\mathbb{E}_2)$ and releases the photon $\hbar\omega_1$, takes time $T_1$. As shown in Fig.11, that is, they located above $x_1$ and  $x_2$ axes. The double arrow straight line on the left between $x_1=0$ and $x_2=0$, it indicates that the electron between the lattice points of the two lines transitions and absorbs the photon $\hbar\omega_1$ with a probability of $1/3$. That is, in the resonant state, the electron transitions from the intermediate energy level $(\mathbb{E}_2)$ to the lowest energy level $(\mathbb{E}_1)$ again. The time taken is $T_2$, as shown below the $x_1$ and $x_2$ axis in Fig.11, and the double arrow straight line on the right between $x_1=0$ and $x_2=0$,  and indicates that the electron between the atomic lattice points of the two lines transitions and absorbs the photon $\hbar\omega_2$ with 1/3 probability. In time $T_1$ or $T_2$, the absorption and release of photons by electrons between $x_1=0$ and $x_2=0$ atomic lattices, and it can be regarded as the exchange of quantum states.

\begin{figure}[t]
	\centering
	{ \includegraphics[scale=0.5]{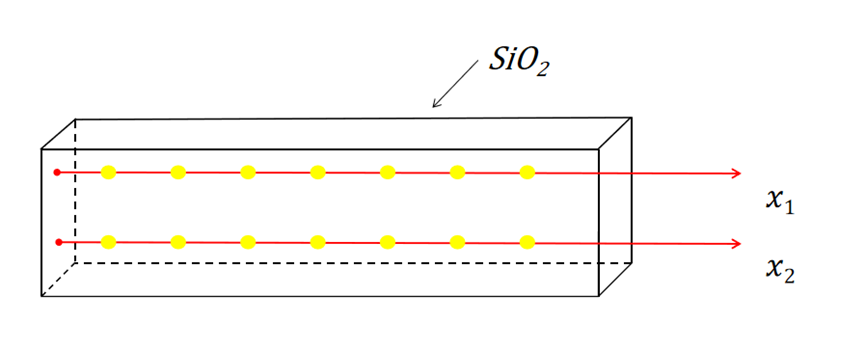}}
	\caption{Two parallel beams passing through a quartz crystal. The $x_1$ expresses photon $\hbar\omega_1$ propagates along the x-axis. In the same way, $x_2$  represents photon $\hbar\omega_2$ propagates along the x-axis.}
	\label{fig:figure 10}
\end{figure}
\begin{figure}[t]
	\centering
	{ \includegraphics[scale=0.9]{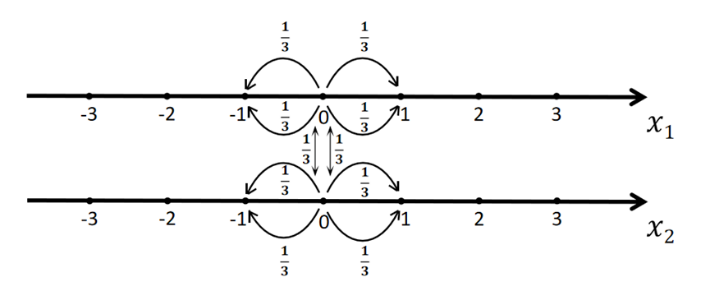}}
	\caption{The first step of quantum walking of four photons on two parallel lines.}
	\label{fig:figure11}
\end{figure}
\begin{figure}[t]
	\centering
	{ \includegraphics[scale=0.24]{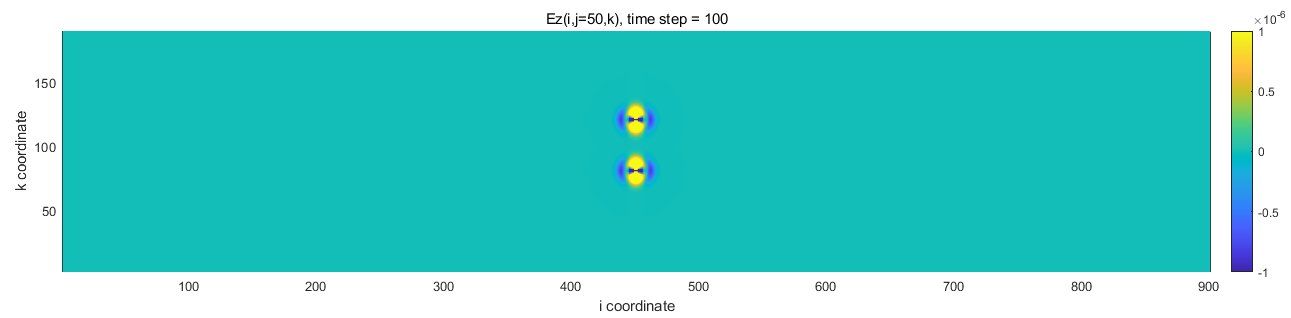}}
	\caption{ Quantum walking of four photons on two parallel lines takes the first step. At $t=T_1$ , the electric field distribution in space is shown.}
	\label{fig:figure 12}
\end{figure}
The electric field distribution corresponding to Fig.11 is shown in Fig.12. At this moment, as can be seen from Fig.12, the electric field is strongly coupled between $x_1=0$ and $x_2=0$. Quantum walking of two parallel line photons takes the second step, as shown in Fig.13. In time $T_1$, the electrons at $x_1=1$ and $x_1=-1$, as well as $x_2=1$ and $x_2=-1$ begin to transition from high energy level to intermediate energy level, releasing the photon $\hbar\omega_1$ with a probability of $ 1/3$, while $x_1=2$ and $x_1=-2$ as well as $x_2=2$  and $x_2=-2$, the electrons absorb photon $\hbar\omega_1$ with a probability of $1/9$ respectively. At this time $t=T_1$, the electrons at $x_1=1$ and $x_1=-1$ as well as $x_2=1$ and $x_2=-1$, absorb the photon  $\hbar\omega_1$ released by the electron transition at $x_1=0$ and $x_2=0$ with a probability of $1/9$, that is, at this time, the photons are exchanged with the nearest atomic lattice around $x_1=0$ and $x_2=0$ with a probability of $1/9$, that is, the exchange of quantum states. At this point, the atomic wave function can be written as:
\begin{equation}
\psi_{x_1=0}=\sqrt{\frac{3}{9}}|\mathbb{E}_3\rangle
\end{equation}
\begin{equation}
\psi_{x_1=1}=\psi_{x_1=-1}=\sqrt{\frac{2}{9}}|\mathbb{E}_3\rangle
\end{equation}
\begin{equation}
\psi_{x_1=2}=\psi_{x_1=-2}=\sqrt{\frac{1}{9}}|\mathbb{E}_3\rangle
\end{equation}

Similarly, when time t is in the time period $T_2$, quantum walking path of photon $\hbar\omega_2$ is the same as the above.
\begin{figure}[t]
	\centering
	{ \includegraphics[scale=0.5]{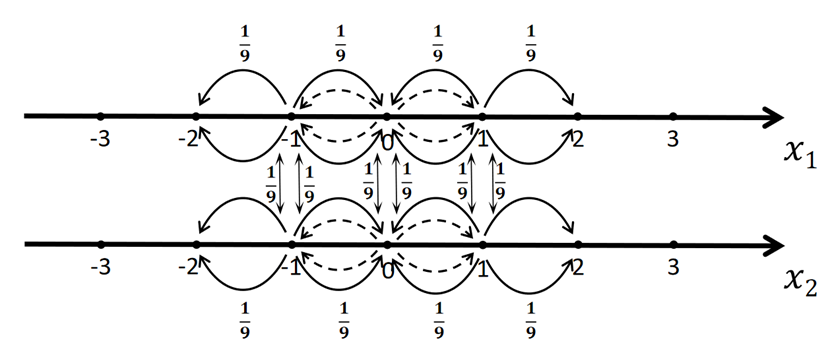}}
	\caption{Quantum walking of four photons on two parallel lines takes the second step.}
	\label{fig:figure 13}
\end{figure}
As can be seen from Fig. 14, the electric field coupling between atomic lattice points is very close, and three excitation sources are generated on each line.
\begin{figure}[t]
	\centering
	{ \includegraphics[scale=0.4]{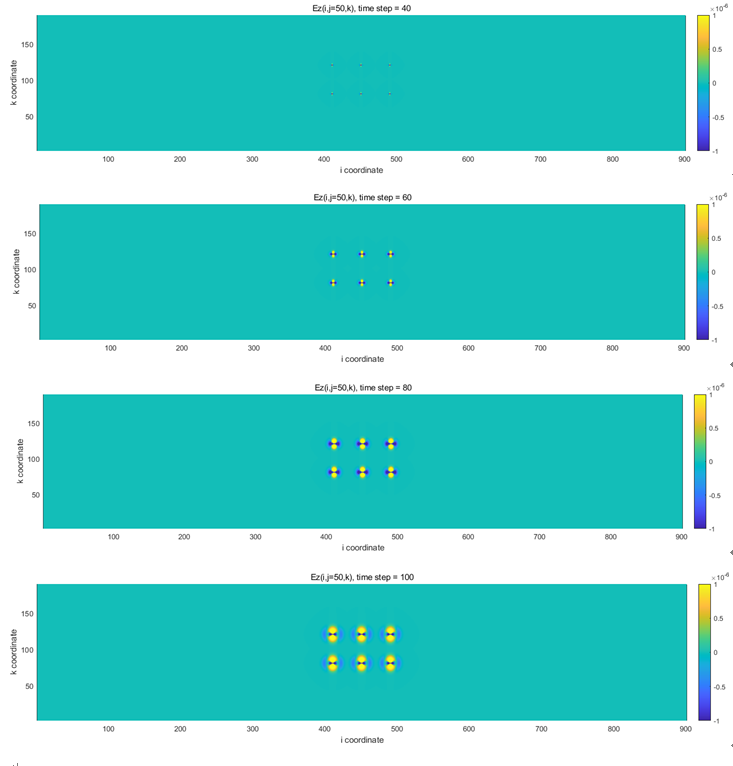}}
	\caption{In the second step of the quantum walking of four photons on two parallel lines, the distribution of electric field in space is from top to bottom, and the time step is from $ndt = 40dt$ to $ ndt = 100dt$. The evolution process of electric field with time is shown.}
	\label{fig:figure 14}
\end{figure}
\begin{figure}[t]
	\centering
	{ \includegraphics[scale=0.5]{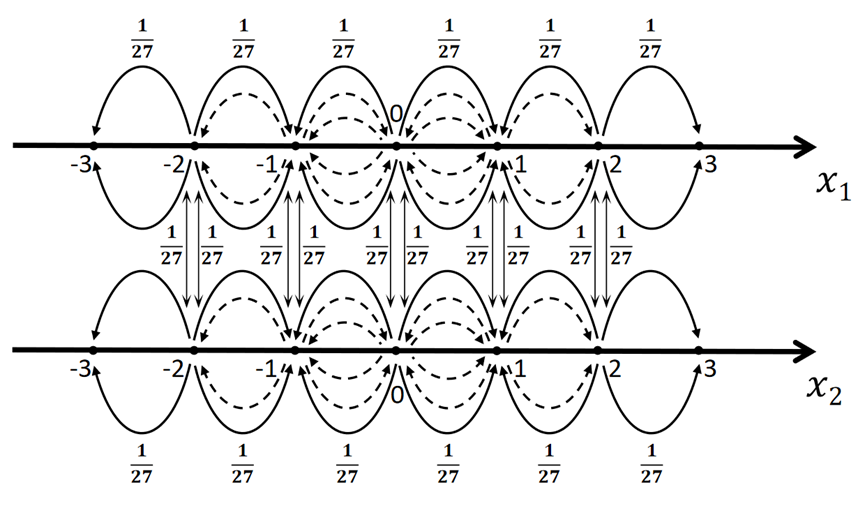}}
	\caption{ Quantum walking of four photons on two parallel lines takes the third step.}
	\label{fig:figure 15}
\end{figure}
As shown in Fig.15, the electrons on the lattice points of $x_1=3$  and $x_1=-3$ as well as $x_2=3$ and $x_2=-3$ absorb the photon $\hbar\omega_1$ with a probability of $1/27$, while the electrons on the lattice points of $x_1=2$ and $x_1=-2$ as well as $X_2=2$ and $X_2=-2$, transition from high energy level to intermediate energy level with a probability of $1/9$, releasing the photon $\hbar\omega_1$. The electrons on the lattice points of $x_1=1$  and $x_1=-1$ as well as $x_2=1$ and $x_2=-1$, absorb photon $\hbar\omega_1$ with the probability of $(1/9+1/27)$ and release photon $\hbar\omega_1$ with the probability of $1/9$, that is, the quantum states with the probability of $1/9$ exchange with each other. Finally, the electrons on the lattice point at $x_1=0$ and $x_2=0$, release the photon $\hbar\omega_1$ from the high energy level to the intermediate energy level with a probability of$1/9$. At the same time, they exchange photons with the electrons on the nearest lattice point with a probability of $1/9$, that is, quantum exchange. The electric field distribution and its evolution with time are shown in Fig.16. The evolution process of electric field with a time $T_1$ .
As can be seen from Fig.16, the coupling of electric field between grid points begins to weaken and spread along the straight line, and five excitation sources are generated on each line. At this time, the wave function on the atomic lattice point can be written as\\
\begin{equation}
\psi_{x_1=0}=\frac{1}{3}|\mathbb{E}_3\rangle
\end{equation}
\begin{equation}
\psi_{x_1=1}=\psi_{x_1=-1}=\sqrt{\frac{1}{9}+\frac{1}{27}}|\mathbb{E}_3\rangle
\end{equation}
\begin{equation}
\psi_{x_1=2}=\psi_{x_1=-2}=\sqrt{\frac{2}{27}}|\mathbb{E}_3\rangle
\end{equation}
\begin{equation}
\psi_{x_1=3}=\psi_{x_1=-3}=\sqrt{\frac{1}{27}}|\mathbb{E}_3\rangle
\end{equation}\\
Similarly, when time  $t$ is in the time period $T_2$, the quantum walks of photons $ \hbar\omega_2$ is the same as the above.\\
In the above discusstion, we suppose that, two quantized harmonic oscillator modes interact with three energy levels, these modes intersect with only one quantized harminic oscillator every two energy levels, and there is only one direct interaction between two of the three possible energy levels. 
Only for a quantum walk on a straight line, its physical process is similar to the traditional discrete-time quantum walk\cite{27}. However, the quantum walk based on two parallel lines has rich physical content. As can be seen from Fig.14 and Fig.16, the third step of quantum walking produces two pairs of excitation sources than the second step of quantum walking, and the excitation sources between the two lines interact with each other and produce the exchange of quantum states. In addition, the time evolution reflects the radiation process of light pulse, that is, from weak to strong, and finally disappear. At the same time, it also shows the instantaneity and ultrashort distance of electromagnetic field.
\section{Conclusions}
Based on the three-level atomic structure and the geometric control theory of quantum computing, this paper gives the equal probability model of quantum walking on a straight line in the resonant state of the system. Through the quantum finite-difference time-domain method, the spatial electric field distribution generated by the first step, the second step and the third step of quantum walking are calculated respectively, and the physical process of each step and the mathematical expression of atomic wave function are given. Next, we give the physical evolution process of the quantum walk on two parallel lines under the condition of synchronization at the same frequency. Using the quantum electromagnetic calculation tool FDTD, the coupling of quantum evolution at a time can be obtained. However, the electric field at this time is instantaneous and ultra-short distance. Moreover, the degree of electromagnetic coupling or electromagnetic coherence between the two wires and the exchange of quantum states can be obtained. The combination of quantum walking and quantum finite-difference time-domain method brings convenience to quantum information processing and provides a practical tool for the development of quantum devices.\\

\begin{figure}[t]
	\centering
	{ \includegraphics[scale=0.4]{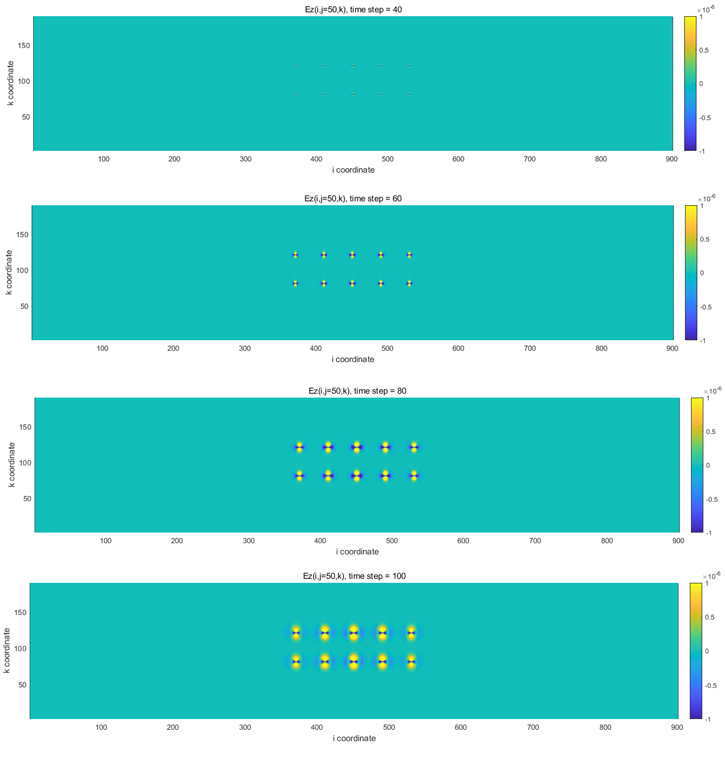}}
	\caption{In the third step of the quantum walking of four photons on two parallel lines, the distribution of electric field in space is from top to bottom, and the time step is from $ndt = 40dt$ to $ ndt = 100dt$.}
	\label{fig:figure 16}
\end{figure}

\bibliographystyle{plain}

\onecolumn\newpage
\appendix

\end{document}